\documentclass[12pt]{article}
\begin{document}
\title{Galaxy Formation with Dark Matter and Dark Energy}
\author{David E. Rosenberg, Manhattanville College Purchase NY  10577\\ \small{email: Rosenbergd1@gradmail.mville.edu}\\}
\maketitle
\copyright 2012
\begin{abstract}

Elliptical and bulge galaxies share a tight correlation of velocity distribution
to both luminosity and black hole mass. There are similar
orbital speeds for all galaxies of a given luminosity including dark matter
(DM) at large radii. The halo surface density of DM is constant
for almost all types of galaxies and ranges 14 mag. down to dwarf
spherical galaxies. There are supermassive black holes or giant, pure
disk galaxies at high redshift inexplicable with hierarchical clustering
or collapse dynamics. These and a myriad of other galaxy
formation problems are explainable by an initial shell which caused 
the Planck cosmic microwave background radiation. 
A reduction in the energy-density of primordial galactic
black holes is necessary to explain dark energy.

\end{abstract}

\newpage

General relativity has been most successful in explaining the universe except in 
extreme density conditions: the big bang and black holes. Among the big bang problems 
is the generating massive galactic structure with an entirely hot model in the first 
billion years. Galaxies are constructed similarly despite their origins being physically
too far apart to be in causal contact (the horizon problem). Initial spacetime was 
Minkowskian, that is the expansion energy exactly matched the gravitational energy  
(the flatness problem). The hot synthesis of the light elements of hydrogen, helium 
and their isotopes occurred in only about 4\% of the initial matter present. 
Extrapolation of general relativity in black holes has resulted in infinities in the 
density and gravity in these structures (singularities). Here is shown 
much evidence that there is a loss of energy at supranuclear densities. 
This will allow a cold shell with a hot core to form. The emitted light will have an 
appropriate (Planck) spectrum, galactic structure will result in the early universe 
and this loss of energy will later be manifested in galactic black holes as dark energy.   

\section{Galaxy Formation Properties and Problems}
Although the $\Lambda$CDM model is quite successful on supragalactic
scales[1], its predictions of galactic properties differ
markedly with observation.
Galaxies come in two basic types: spirals which are disk shaped and
ellipticals which are football shaped. The accepted theory is that the
present universe grew from small inhomogeneities. These grew into larger
halo structures by attracting surrounding matter. The fate of these halos
is determined either by radiative cooling or gravitational heating. In low
mass halos, cooling predominates, which allows cold gas to fall into the
center and become disks and stars[2]. The cooling problem is
most acute in galaxies. At the end of their lives, massive stars return
$30-40\%$  of their mass to the interstellar media. If even a small
fraction of this mass is accreted, it would result in much larger black
holes than are present. Gravitational heating dominates once a halo mass of about $10^{12}M_\odot$
is reached. Cold gas is no longer able to accrete onto galaxies.
The only way galaxies within halos can grow at this stage is by mergers.
Pure disk galaxies form bulges after the mergers. Yet some samples of  giant galaxies
have found over half are large pure disk type, without any evidence of
mergers[3].

In galaxies with bulges, the mass of the central black hole correlates with
the mass of the bulge and also the average spread of velocities of the bulge
stars. This includes ellipticals which have bulges but no disks[4].
The mechanisms for these correlations are not known.
Another problem is why the gas that formed bulge stars settled
near the black hole. Part increased the black hole mass and part led to
explosions that blew the gas away and suppressed star formation. Many
components of galaxies besides black holes are highly correlated.
The mass distribution of spiral galaxies is evenly spread from its dark
matter outer limits to its inner baryonic areas. Dark matter played a
strong role in the disk and stars but not its black hole. In pure disk
galaxies with pseudobulges, the central black hole does not correlate with
the pseudobulge[5]. Another puzzle is the reason for the inward
movement of matter to the black hole in some galaxies and to the pseudobulge in
others. In accepted galaxy formation theory, both galaxies with and without bulges
grew by accreting matter during the period that the massive early stars were
forming. These early stars would not have settled in disks because they could
not be slowed enough to reside in disks. Galactic bulges do contain old stars
but there is no reason these old stars avoided bulgeless galaxies.
There is no evidence that they are in diffuse stellar halos
either[4]. Evidently they played little role in galaxy formation. 
The DM halo mass distribution for galactic systems ranging from dwarf discs
and spheroidals to spirals and ellipticals is essentially constant[6]. This
amazing result also spans almost the whole galaxy magnitude range $M_B$ from
$-8$ to $ -22$ and gaseous to stellar mass fraction range of many orders of magnitude.
\begin{equation}
log\frac{\mu_{0D}}{M_\odot pc^{-2}}=2.15 \pm 0.25
\end{equation}
where $\mu_{0D}$ is the central surface density and is defined as 
$r_0\rho_0$. $r_0$ is the halo core radius and $\rho_0$ is the central density.
 
Simulations of galaxy formation usually start
with a set of hot gravitating point particles with given 
initial conditions which are then stepped forward
in time using huge computer resources. The Jeans mass is thought to
be the point at which gravity overcomes pressure and expansion to
form galactic structure. Collapse dynamics produces different post
collapse densities, circular speeds and disk asymmetries. Both collapse 
and hierarchical clustering approaches have
been unable to solve the many problems which are summarized here from[7]. 
1. Is the absence of a feature in galaxy rotation curves at which the dominant source
of central attraction changes from luminous matter to dark. Many galaxies
are now known in which the rotation curve does drop somewhat at
the edge of the visible disk, but it is extremely rare for the drop to exceed about
$10\%$. A featureless rotation curve is
expected if DM dominates galaxies right to their centers, but it is much
harder to understand why the circular orbital speed from the luminous
matter, which dominates the inner region, should be so similar to
that from the DM at larger radii. For any galaxy dominated by stars in its
center, initial conditions for the dark and luminous matter must be finely
tuned to produce a flat rotation curve. 
2. Extreme low-SB galaxies lie on the same Tully-Fisher relation (TFR) derived
from high-SB galaxies, with somewhat greater scatter. There are
similar circular speeds in all galaxies of a given luminosity, no matter how
widely the luminous material is spread. This unusual result requires that
the overall mass to light ratio (M/L) of the galaxy rises with decreasing SB in just the right
way so as to preserve a tight relation between total luminosity and circular
speed. Either the true M/L of the stellar population changes with surface
brightness, which seems unlikely, or the DM
fraction rises as the luminous surface density declines. The needed variations
would be minor if DM dominated in all galaxies, but since stars
dominate the mass in the inner parts of high-SB galaxies, eliminating
the SB dependence again requires careful tuning.
3. Mass discrepancies begin to be detectable only when the acceleration drops below
$\sim 10^{-8} cm/s^{2}$. Any DM model must reproduce this characteristic
acceleration scale over a wide range of galaxy sizes.
4. Aside from the disk M/L, DM halo fits to rotation curves generally employ
two extra parameters: e.g. the core radius and asymptotic velocity, or the
scale radius and concentration index. Actual galaxy rotation curves do not
require all this freedom, however, since they can be fitted with only the
disk M/L as a parameter. Any DM model must therefore contain a physical 
mechanism that relates the halo parameters to the luminous mass distribution.
5. A merging hierarchy causes the cooled baryonic fraction to lose angular
momentum to the halo, making disks that are too small. The predicted 
angular momentum of the disk is at least an order of magnitude less than 
that observed. The problem is only partially ameliorated if some process 
(usually described as "feedback from star
formation") prevents most of the gas from cooling until after the galaxy is
assembled. While this difficulty is best known within the CDM context,
merging protogalaxies in any hierarchical structure formation model with
DM will involve chronic angular momentum loss to the halos.
6. Every collapsed halo should manifest the same peak phase space density,
 if DM is collisionless, was initially homogeneously distributed, and
had an initially finite value (Liouville's theorem). Further, the finite central
density of galaxy halos also suggests that halos collapsed from material
having an initially finite value (infinite initial phase space density forms
cusped halos) allows it to be estimated easily. Its spectacular
variation between galaxies indicates that DM cannot be a simple collisionless
particle.
7. High-resolution simulations that follow the formation and evolution of individual
galaxy halos in CDM find strongly cusped density profiles even before the baryonic 
component cools and settles to the center. No observational evidence requires halos to have
the predicted cusps. Further, the "concentration index" has a wide range
but most fits to rotation curves yield values well below the predicted range 
in all types of galaxy, even the Milky Way.
8. There is a failure to predict the zeropoint of the Tully-Fisher relation (TFR). 
This is a major problem for the $\Lambda$CDM paradigm." No matter what M/L is assumed for the disk, 
the predicted circular speed at a given luminosity is too high because the halo density
is too high.
9. Simulations produce numerous sub-clumps within large DM halos. 
The clumps are more numerous than the
numbers of observed satellite galaxies, and may threaten the survival of a
thin disk in the host galaxy.
10. The TFR discrepancy is even worse, since CDM predicts $L \propto V^3$. 
If V is interpreted as the circular velocity of the
flat part of the rotation curve, the true relation is very nearly $L \propto V^4$.
Any mechanism which systematically boosts luminosity as a function of
mass must also reproduce the very small scatter in the TFR.
11. The DM halos that form in simulations are generally tri-axial, 
but become nearly oblate in their inner
parts when a disk is added. Current constraints on halo
shapes are generally thought to be consistent with these
predictions. However, the halo of NGC 2403 seems to become more nearly
axisymmetric at larger radii, opposite to the CDM
expectation, and IC 2006 seems impressively round at $6R_e$. 
12. The CMBR problem that was originally listed has changed as more accurate analyses 
have emerged. There is another problem which seems most difficult for both the  
hierarchical merger and collapse theories of galaxy formation. 
This is the finding of a galaxy with a supermassive black hole
of two billion solar masses at a redshift  $z=7.085$, 
just $770$ million years after the big bang[8]. Finding supermassive black
holes in galaxies at high redshift have always been a problem for accepted galaxy formation 
methods and totally hot big bang models, but a finding this early makes it acute.

\section{Galaxy formation with primordial black holes}
A shell of baryonic matter can supply the seeds to
initiate massive black holes and capture hot expanding core gases along with 
smaller shell matter for the haloes. The deeper the gravitational 
wells, the higher the velocity and more orbiting mass that could 
be captured. The capturing process described here is divided
by distance from the primordial black holes.

Outside the immediate area of black hole influence, capturing
of hot core matter streaming through the area of influence of each black hole
is due to the amount of energy each particle possesses. Large
kinetic energies result in hyperbolic or parabolic type orbits
with the ability to escape any given gravitational well. Lower energies
result in stable elliptical or circular orbits.
\begin{equation}
e=\sqrt{1+\frac{2El^2}{mk^2}}
\end{equation}
where E is the total energy, both kinetic and potential. 
$l$ is the angular momentum, $M$ is the central black hole mass and
and $m$ is the rotational mass. If $e<1$ and
$E<0$, the orbit is an ellipse and the matter
will be captured. Circular orbits where $e=0$ and 
$E=-{mk^2}/{2l^2}$ have even less energy.
Matter that is captured has the potential
energy greater than the kinetic, 
\begin{equation}
\frac{GmM}{r}>\frac{l^2}{mr^2}+\frac{1}{2}m\dot r^2
\end{equation}
and $e<1$. Expanding the total kinetic energy
$E$ in the equation for $e$,
\begin{equation}
e=\sqrt{1+\frac{2l^2(l^2/mr^2+\frac{1}{2}m\dot r^2-GmM/r)}{mk^2}}
\end{equation}
Orbiting matter has $e<1$ and real. If we let the angular momentum
$l=mr\dot \theta^2$ and $k=mMG$, the equation for $e$ becomes
\begin{equation}
e=\sqrt{1+\frac{r^6\dot\theta^4+\dot r^2r^4\dot\theta^2
-2GMr^3\dot\theta^2}{M^2G^2}}
\end{equation}
To simplify this equation, we can use $\dot\theta=\dot r/r$.
The equation for $e$ becomes
\begin{equation}
e=\sqrt{1+\frac{2r^2\dot r^4}{M^2G^2}-\frac{2r\dot r^2}{MG}}
\end{equation}
As $GM=\dot r^2r$, 
then the galactic well will deepen as $M\propto \dot r^2$
or $M\propto r$. The last term in equation above becomes $\dot r^8/M^2G^2$. When this
term is dominant, it will allow capturing matter with $\dot r$ to 
increase as the fourth power as the galactic black hole $M$ increases,
$\dot r \propto M^4$. This explains the Tully-Fisher and Ferrarese-Merritt[9] relations. 
The black hole capturing cross sectional area, 
$M_{csa}\propto M_{gravity}$ since both scale as $r^2$. Other factors during 
galaxy formation may include 
accretion due to collisional losses, ionized plasma repulsion and magnetic interference.

Effective potential for motion in Schwarzschild geometry[10] 
with a mass $M$,
energy in units of rest mass $\mu$ of the particle is $\tilde{E}=E/\mu$
and angular momentum is $\tilde{L}=L/\mu$. The quantity $r$ in the next
equations is the Schwarzschild coordinate.
\begin{equation}
\bigr ({dr\over d\tau} \bigl )^2 + \tilde{V}^2(r)=\tilde{E}^2 
\end{equation}
and therefore  
\begin{equation}
\tilde{V}^2=(1-2M/R)(1+\tilde{L}^2/r^2)
\end{equation}
Stable orbits are possible for $\tilde{L}>2\sqrt{3}M$.  
The above equations are for nonrotating or slowly 
rotating black holes. For an unbound orbit, the impact parameter $b$ is
\begin{equation}
b=\tilde{L}/\sqrt{(\tilde{E}^2-1)}
\end{equation}
The capturing cross section for a nonrelativistic particle
\begin{equation}
\sigma_{capt}=16\pi M^2/\beta^2
\end{equation}
where $\beta$ is the velocity relative to light. For relativistic particles 
\begin{equation}
\sigma_{capt}=27\pi M^2(1+\frac{2}{3\tilde{E}^2})
\end{equation}

\section{Changes In Our Universe}
A cold baryonic shell surrounding a hot core can absorb and not reflect hot core photons. 
It comprised a cavity close to the characteristic of a perfect black body with resulting radiation 
in thermal equilibrium as shown in Figure~1. Whether released from a small hole or the massive break 
up of the shell, subsequent light emission would have a Planck spectrum, like the CMBR. 
The emitted light was homogeneous, isotropic, unpolarized and had power emitted at an angle 
to the normal, proportional to the projected area. The spectral energy density was 
\begin{equation}
u_\nu (T)= \Bigr (\frac{8\pi\nu^2}{c^3}\Bigl ) \Bigr (\frac{h\nu}{e^{h\nu/k_BT}-1}\Bigl )
\end{equation}
where $k_B$ is Boltzmann's constant. The first term on the right represents 
the number of electromagnetic modes of the standing waves at frequency $\nu$ 
per volume of cavity. The second term represents
the average energy per mode at this frequency. 
For a derivation of Planck cavity radiation, see[11]. 
This model is fully consistent with the latest seven year CMBR study by WMAP[12],
including primordial power spectum and Gaussianity, neutrino properties, helium abundance,
parity violation, dark energy, and polarization. Only the dark matter are cold baryons. 

The finding of a constant DM halo surface density across all types and sizes of
galaxies confirms that this part must have been the first to form.
As the big bang unfolded, the shell ruptured as fragments, large and small, were propelled into the universe.
Blacks holes were formed from the larger parts. 
As the higher velocity hot core matter followed, angular momentum was preserved
by the gravitational capturing process. 
Black holes form at densities
\begin{equation}
\rho = \frac{c^6}{G^3M^2}
\end{equation}
The smallest black holes that have been found in the present universe 
are all above $4.3 M_\odot$, leaving a gap above the most massive neutron stars[13]. 
There must be a limiting supranuclear mass density of $\rho \approx 2.4 \times 10^{16} g/ cm^3$. 
After all the space is squeezed out of nucleons, it must cost energy to collapse further.
This possibility was discussed in Misner, Thorne and Wheeler[10], the Bible on gravitation, 
page 627. Our galaxy shows evidence of this energy and gravitational loss tunneling out of 
the central black hole after billions of years. There is a lopsided thickness in the outer disk of the Milky Way, 
which is about twice as large in one side as the other, despite the spherically 
symmetric distribution of galactic dark matter[14]. The tunneling of particle waves
and the resulting net loss of energy and gravity is shown in Figure~2.
This late energy loss can be seen as well in the measured supernova distances. 
The 3-geometry $d\sigma^2= g_{ij}(t,x^t)dx^idx^j$ of each hyperspace is expected to be the same
due to homogeneity of the universe. The initial hypersurface $S_I$ 3-geometry is
$\gamma_{ij}x^K \equiv g_{ij}(t_I,x^K)$. At time $t_I$ on surface $S_I$, they are separated by
the proper distance $\Delta\sigma(t_I)=(\gamma_i,\Delta x^i \Delta x^j)^{1/2}$. At some later time
$t_f$, they will be separated by some other proper distance $\Delta\sigma(t_f)$. When spacetime is
isotropic, then the ratio of $\Delta\sigma(t_f)/\Delta\sigma(t_I)$ will be related to the universe expansion 
$a(t_f)/a(t_I)$. The loss in energy and gravitation will cause galaxies to wander off the Hubble flow with 
greater distances measured between them and is 
known as dark energy. This is demonstrated in Figure~3. 
This density limitation is also involved during supernova collapses. 
Although the large production of neutrinos will slow or stop a shock front, an incompressible 
core is necessary to restart it as a bounce.
 To calculate the standard electromagnetic energy density 
\begin{equation}
\epsilon=\frac{1}{8\pi}(E^2+B^2), S=\frac{1}{4\pi}(E\times B)
\end{equation}
where $\epsilon$ is the total density of mass-energy, S is the total flux of mass-energy or momentum
density and B and E are magnetic and electric fields, respectively. The stress-energy tensor is
\begin{equation}
T_{jk}=\frac{1}{4\pi}[-(E_jE_k+B_jB_k)+1/2(E^2+B^2)g_{jk}]
\end{equation}
or eqivalently
\begin{equation}
T_{jk}=(\rho_m+p)\gamma^2v_jv_k+pg_{jk}, \gamma=(1-v^2)^{-1/2}
\end{equation}
As core density reached $\rho \ge 2.4 \times 10^{16}$ space was eliminated from the nucleons. 
As the collapse proceeded, energy was stored in core quark compression, an energy sink. 
Finally the entire structure's gravitational energy was concentrated in the core.
This stored potential energy was released as kinetic energy when the shell broke up,
allowing nucleosynthesis only in the core matter. 
The energy sink caused the mass-energy of the entire structure to be zero. 
With this change, energy and momentum remain balanced so that the divergence $\mathbf{\nabla \cdot T}=0$. 
In black holes the lapse (of time) function, black hole proper time to universal time, 
$\alpha \equiv d\tau/dt$ is thought to be zero[15]. 
Using the dark energy time of manifestation in our universe, $\alpha \sim 2 \times 10^{-16}$, small but not zero.
Supporting the density limitation, neither small black holes less than $4M_{\odot}$ nor Hawking radiation has been found.
The stress-energy tensor $T_{jk}$ will remain a second order symmetric tensor like the Einstein curvature tensor $G_{jk}$, 
and both will be zero until the breakup of the shell.
The universe did not become approximately isotropic and homogeneous until hours after the big bang
as the hot core gases were captured by the black holes forming protogalaxies. 
Only after this period will Alexander Friedmann's equations with 
time derivatives of the universe scale factor become valid.
With a density limitation in the stress-energy tensor, 
general relativity can be valid from the very beginning of the universe and throughout black holes. 

Halo parameters are related to the luminous mass distribution since all
rotating mass was captured by a given size black hole as shown in Figure~4.  
An entirely baryonic model explains why the circular orbital speed 
from luminous matter, which dominates the inner regions, is so similar
to dark matter at larger radii. With many stars in the center areas, initial
conditions for dark and luminous matter no longer have to be closely adjusted 
to produce a flat rotation curve. 
A core based expansion can be captured by the similarly sized black holes, 
explaining why there are similiar circular speeds in 
all galaxies of a given luminosity no matter how the luminous matter is spaced.
The overall mass to light ratio rises with decreasing surface brightness so as
to preserve the Tully-Fisher relation between total luminosity and circular
speed. The depth of the gravitational well determines the circular speed
and luminosity. The hot and cold matter discrepancies are detectable only at
accelerations below $\sim 10^{-8}cm/sec^{2}$ since they are all baryons. 
Collapse dynamics seems unnecessary in 
initial galaxy formation except possibly in pseudobulge galaxies. 
The peak phase space density of the halo varies so markedly as 
baryonic dark matter is not collisionless and was never 
homogeneously distributed. Both shell and core baryons are included. 
The predicted circular speed at a given luminosity is high. 
The capturing of slower velocity matter allowed an increase in kinetic energy
as it fell into the gravitational well. It is not directly related to 
mass to light ratio or halo density. The number of predicted subclumps in the
halos is so much greater than observed as well as predicted clumps 
compared to satellite galaxies because of the capturing process rather than collapse dyamics. 
As shown above, the luminosity can be related to circular velocity to the fourth power. 
Primordial massive black holes can efficiently remove intergalactic and intercluster
matter in the early universe. 

\section{Discussion}
General relativity has been extrapolated in black holes and the big bang to 
enormous energies and densities without sufficient supporting data. 
Using evidence of a limitation in the stress-energy tensor, a viable
big bang model can be produced, early and accurate galaxy formation can be obtained, 
black holes can lose information, and dark matter and dark energy can be explained. This 
should remove the conflict between quantum physics and general relativity and greatly
simplify a theory of everything.

\section{Bibliography}

1. Bahcall, N. et al. The Cosmic Triangle: Assessing the State of the 
Universe. Science 284, 1481-1488 (1999).

2. Cattaneo, A. et al. The Role of Black Holes in
Galaxy Formation and Evolution. Nature 460, 213-219 (2009).

3. Kormendy, J. et al. Bulgeless Giant Galaxies Challange Our Picture of 
Galaxy Formation by Hierarchical Clustering. preprint at http: 
//arxiv.org/abs/1009.3015 (2010).

4. Peebles, P. J. E. How Galaxies Got Their Black Holes. Nature 469, 305-306 (2011).

5. Kormendy, J., Bender, R., \& Cornell, M.E. Supermassive Black Holes
Do Not Correlate with Galaxy Disks or Pseudobulges. Nature 469, 374-376 (2011).

6. Donato, F. et al. A Constant Dark Matter Halo Surface Densities in Galaxies. 
Mon. Not. R. Astron. Soc. 397, 1169-1176 (2009).

7. Sellwood, J.A. \& Kosowsky A. Does Dark Matter Exist? Gas and Galaxy
Evolution, ASP Conference Series (2000).

8. Willott, C. A Monster in the Early Universe. Nature 474, 583-584 (2011).

9. Merritt, D. \& Ferrarese, L. Supermassive Black Holes. Phys. World 15N6, 41-46 (2002). 

10. Misner, C., Thorne, K.S., \& Wheeler, J.A. Gravitation  (W.H. Freeman and Co., New York, 1973).

11. Brehm, J. J. \& Mullin, W.J. Introduction To The Structure Of Matter 
 (John Wiley and Sons, New York, 1989) 73-99. 

12. Komatsu, E. et al. Seven Year Wilkinson Microwave Anisotropy Probe (WMAP): 
Cosmological Observations. Astrophys. J. Suppl. 192, 18 (2011).

13. Farr, W. M. et al. The Mass Distribution of Stellar Mass Black Holes. 
Astrophys. J. 741, 103-122 (2011).

14. Blitz, L. The Dark Side of the Milky Way. Scientific American 305,4, 36-45 (2011).

15. Thorne, K.S., Price, R.H., \& Macdonald, D.A. Black Holes: The Membrane
Paradigm (Yale University Press, New Haven, 1986).

\textbf{Acknowledgements} This article is based on research posted on the internet 
astrophysics archives. Some of it was done in collaboration with 
John Rollino of Rutgers University Physics, Newark N.J. 07102

\textbf{Author Information} Reprints and permissions information is available at 
www.nature.com/reprints. The author declares no competeing financial interests. Readers
are welcome to comment on the online version of the article at www.nature.com/nature.
Correspondence and request for materials should be 
addressed to rosenbergd1@gradmail.mville.edu

\end{document}